\documentclass[pre,aps,superscriptaddress,twocolumn,amsmath,amssymb,floatfix,10pt]{revtex4-2}
\usepackage[breaklinks,colorlinks = true,linkcolor = magenta,urlcolor=magenta,citecolor=red,hypertexnames=false]{hyperref}
\usepackage[section]{placeins} 
\usepackage{xfrac} 
\usepackage{float} 
\usepackage{svg} 
\usepackage{bm} 

\usepackage{soul} 

\graphicspath{{./}{./figures/}}
\allowdisplaybreaks

\begin{document}

\title{Coarse grained descriptions of the dynamics of yielding of amorphous solids under cyclic shear}

\author{Debargha Sarkar}
\affiliation{Jawaharlal Nehru Centre for Advanced Scientific Research, Jakkur Campus, Bengaluru, 560064, India}

\author{Jishnu N. Nampoothiri}
\affiliation{Jawaharlal Nehru Centre for Advanced Scientific Research, Jakkur Campus, Bengaluru, 560064, India}

\author{Muhittin Mungan}
\affiliation{Institute for Biological Physics, University of Cologne, Cologne, Germany}

\author{Jack T. Parley}
\affiliation{Institut für Theoretische Physik, University of Göttingen, Friedrich-Hund-Platz 1, 37077 Göttingen, Germany}

\author{Peter Sollich}
\affiliation{Institut für Theoretische Physik, University of Göttingen, Friedrich-Hund-Platz 1, 37077 Göttingen, Germany}
\affiliation{Department of Mathematics, King’s College London, London WC2R 2LS, UK}

\author{Srikanth Sastry}
\affiliation{Jawaharlal Nehru Centre for Advanced Scientific Research, Jakkur Campus, Bengaluru, 560064, India}

\begin{abstract}
Recent computer simulations reveal several intriguing features in the evolution of properties of amorphous solids subjected to repeated cyclic shear deformation. These include the divergence of the number of cycles to reach steady states as the yielding point is approached, a non-monotonic change of properties with cycles, and the possibility of a spectrum of frozen states. Theoretical attempts to capture these properties through simple models, including the Ehrenfest model describing a random walk in a confining potential, have met partial success. Here, we show that incorporating the influence of mechanical noise through a feedback term leads to a genuine dynamical transition with characteristics reflecting those of yielding. Coarse graining the dynamics into a small number of variables leads to new insights regarding the dynamics of yielding.

\end{abstract}
\maketitle

Amorphous solids, such as glasses, are ubiquitous in nature, and, because of their widespread use in engineering and material science applications, their mechanical properties, in particular their yielding behaviour under applied stress/deformation, are of broad interest. Further, the structural disorder of amorphous solids and the out-of-equilibrium nature of external driving leading to yielding, present fundamental challenges to developing a satisfactory theoretical description. Yielding behaviour has been actively investigated in recent years, theoretically and using computer simulations, both under uniform shear and cyclic shear deformation\cite{leishangthem_yielding_2017,ozawa_random_2018,bhaumik_role_2021,shi_evaluation_2007,yeh_glass_2020,vasisht_emergence_2020,popovic_elastoplastic_2018,talamali_strain_2012,barlow_ductile_2020, kumar2022}. 
In particular, several simulation studies focus on athermal quasistatic (AQS) shear deformation, which will provide the context for the discussion to follow \cite{fiocco_oscillatory_2013,kawasaki_macroscopic_2016,leishangthem_yielding_2017,regev_onset_2013,ozawa_random_2018,priezjev_yielding_2018}.

In the case of cyclic shear, wherein shear deformations of a given amplitude are applied repeatedly, amorphous solids undergo irreversible plastic rearrangements leading to an evolution of their properties, till they reach steady states, which are non-diffusive for small amplitudes of shear and diffusive for amplitudes above the yield strain amplitude. The number of cycles to reach the steady states increases as the yielding point is approached on either side, with an apparent power-law divergence \cite{fiocco_oscillatory_2013, regev_onset_2013,leishangthem_yielding_2017,PRIEZJEV2023112230,kurotani2022fatigue,maity2024fatigue} \footnote{Elasto-plastic simulations \cite{liu_fate_2022,khandare20xx}, however, appear to indicate a logarithmic divergence as the yielding point is approached from below.}. For amplitudes larger than the yield value, thus, the number of cycles to failure increases as the deformation amplitude is reduced, diverging at the yield amplitude, which thus may be identified as the  {\it fatigue  limit}\cite{kunFatigueFailureDisordered2007,kun_universality_2008,Basquin}, discussed in the context of fatigue failure of materials.  Computer simulations of model glasses also reveal that their yielding behaviour under cyclic shear is strongly dependent on their initial degree of annealing, as may be measured by their interaction energy \cite{bhaumik_role_2021,bhaumik_yielding_2022,yeh_glass_2020}. Glass samples prepared above a {\it threshold} energy display mechanical annealing up to a common strain amplitude, termed the {\it yield point}, where their energy (measured at zero applied strain)  achieves the threshold value irrespective of the initial condition. For samples prepared below the threshold energy, both the strain amplitude at which yielding occurs, and the magnitude of discontinuity of properties, increases with the degree of annealing (lowering of energy). Thus, glasses below the yield point can be in one among a range of annealing states, and even above the yield point, a subset of these glasses can remain stable and not yield. Above the yield point, glasses that eventually fail display interesting non-monotonic dependence of properties (including their energy, and measures of plastic activity), suggesting the co-existence of processes that lead to annealing and failure. A successful description of cyclic shear yielding needs to provide an explanation and insight into these observations. 

Given the central role of disorder and annealing, energy landscape based models of plasticity and yielding (such as the soft glassy rheology model~\cite{sgr_sollich})  are attractive, and have recently been investigated \cite{sastry_models_2021, parley_mean-field_2022, mungan_metastability_2021}. In these and related works, one models the states that are meant to describe a mesoscopic block of a solid, and transitions between them,  with the effects of the surrounding system incorporated with varying degrees of realism. Qualitative agreement with simulation results was demonstrated for models (which describe a single block) considered in \cite{sastry_models_2021}, which nevertheless do not display a genuine dynamical transition corresponding to yielding. A mean-field theory of cyclic shear yielding of such a model, investigated in \cite{parley_mean-field_2022}, incorporating mechanical noise arising from interactions between blocks, along lines studied for elasto-plastic models \cite{nicolas_deformation_2018,lockwoodUltradelayedMaterialFailure2023, cochran_slow_2024}, does exhibit a dynamical transition, as discussed further later. In \cite{mungan_metastability_2021}, the dynamics of the (single block) models was further mapped, considering the states reached at the end of each cycle of deformation, to a random walk along the energy axis, in the presence of a confining potential, and an absorbing boundary which corresponds to the system reaching states that were stable with respect to a given shear deformation. Thus, the yielding problem is mapped to the well-known Ehrenfest urn model. The possibility of escape from mechanically stable states was also considered, with the source of noise being thermal, leading to a well-defined yield point but no genuine dynamical transition.

In the present work, we consider a variant of the model in \cite{mungan_metastability_2021} and extend it to include activation due to {\it mechanical noise}, arising from plasticity elsewhere in the system. To this end, we first consider the incorporation of mechanical noise in the Ehrenfest model and show that it leads to a genuine dynamical transition, both by numerically simulating the model and from a simplified coarse-graining with the help of which we calculate the form of the divergence of cycles needed to reach the steady states. We next analyse general 
features of the master equation governing the dynamics of the system, and demonstrate that a minimal three-state 
description is sufficient to capture the non-monotonic behaviour of time-dependent properties, as well as to characterise the range of possible frozen states. We finally derive such a three-state description by coarse graining the Ehrenfest model and show that it indeed displays the salient features of the dynamics of yielding under cyclic shear deformation. We also show that near the critical strain amplitude, the dynamics of a broad class of models follows a universal scaling curve.

{\it The Ehrenfest model:}  The models investigated in \cite{sastry_models_2021,parley_mean-field_2022} are based on an energy-landscape picture, and consider a distribution $\rho(\epsilon)$ of possible states for a mesoscopic block in the amorphous solid, and their response to driving {\it via} cyclic shear with amplitude $\gamma$.  Each state is marked by an energy  $\epsilon$ (deeper states have larger $\epsilon$), at a stress-free (or plastic) strain value, $\gamma_0$, and a stability range in strain $\gamma^{\pm}_\epsilon =   \gamma_0 \pm \delta \gamma_{\epsilon} $, with $\delta \gamma_\epsilon\propto \sqrt{\epsilon}$, within which the energy increases quadratically with strain. (For simplicity, we choose $ \gamma_0 = 0$, 
$\epsilon (\gamma) = \gamma^2$, and $\delta \gamma_\epsilon = \sqrt{\epsilon}$ in the following). 
When the applied strain exceeds the stability range, a transition occurs to a new state drawn from the distribution of possible states.  States with sufficiently large $\epsilon$ ($\epsilon>\gamma^2$ for $\gamma_0 = 0$) are stable with respect to successive cycles of shear, whereas unstable states may undergo multiple transitions during a cycle. The dynamics can thus be thought of as a series of transitions between energy states from one cycle to the next, till a stable state is found. We consider here a generalization \footnote{The reference distribution in \cite{mungan_metastability_2021} was binomial, whereas here we chose the distribution to be half-Gaussian, as in \cite{sastry_models_2021}.} of the stochastic model  considered in \cite{mungan_metastability_2021}, with discrete energy states $\epsilon_i=i/2N$ where $i\in\{0,\ldots,2N\}$ and transition probabilities to neighbouring energies $j\to j\pm 1$ given by 
\begin{equation}\label{eq:transition_rules}
p_{j\pm 1,j}=\frac{1}{2}\left[1\pm \frac{1}{4}\ln\left(\frac{\rho(\epsilon_{j+1})}{\rho(\epsilon_{j-1})}\right) \right]\end{equation}
These transition probabilities, together with the rates $1/\tau_j$ (in units of number of cycles) at which transitions out of each state $j$ occur, determine the time-dependent probabilities $p_i(t)$. The latter obey a master equation as detailed in the Supplemental Material (SM) \cite{SMref}  
; physically, they represent the fraction of mesoscopic blocks occupying the different states. 
The assignment (\ref{eq:transition_rules}) of transition probabilities is chosen so that for state-independent $\tau_j=\tau_0$, the $p_i(t)$ approach a steady state probability distribution $q_j\propto \rho(\epsilon_j)$.

To take into account the difference between stable states, i.e.\ those with $\epsilon_j \ge \gamma^2$, and unstable ones, the simplest model in \cite{mungan_metastability_2021} sets $1/\tau_j=0$ for stable states. The stable states are then absorbing and the dynamics is governed solely by the time needed for the system to hit the set of stable states (S) starting from the unstable states (U) (denoted by $W_{SU}^{-1}$, see below), which is a monotonically increasing function of strain amplitude $\gamma$. The system will eventually reach an absorbing state in $\cal{O}$$(W_{SU}^{-1})$ cycles and remain frozen thereafter, regardless of the amplitude of shear. 

A more realistic picture emerges when one considers the possibility that notionally stable states may yield due to the effects of noise.

\begin{figure}[!htbp]
    \centering
    \includegraphics[scale=0.28]{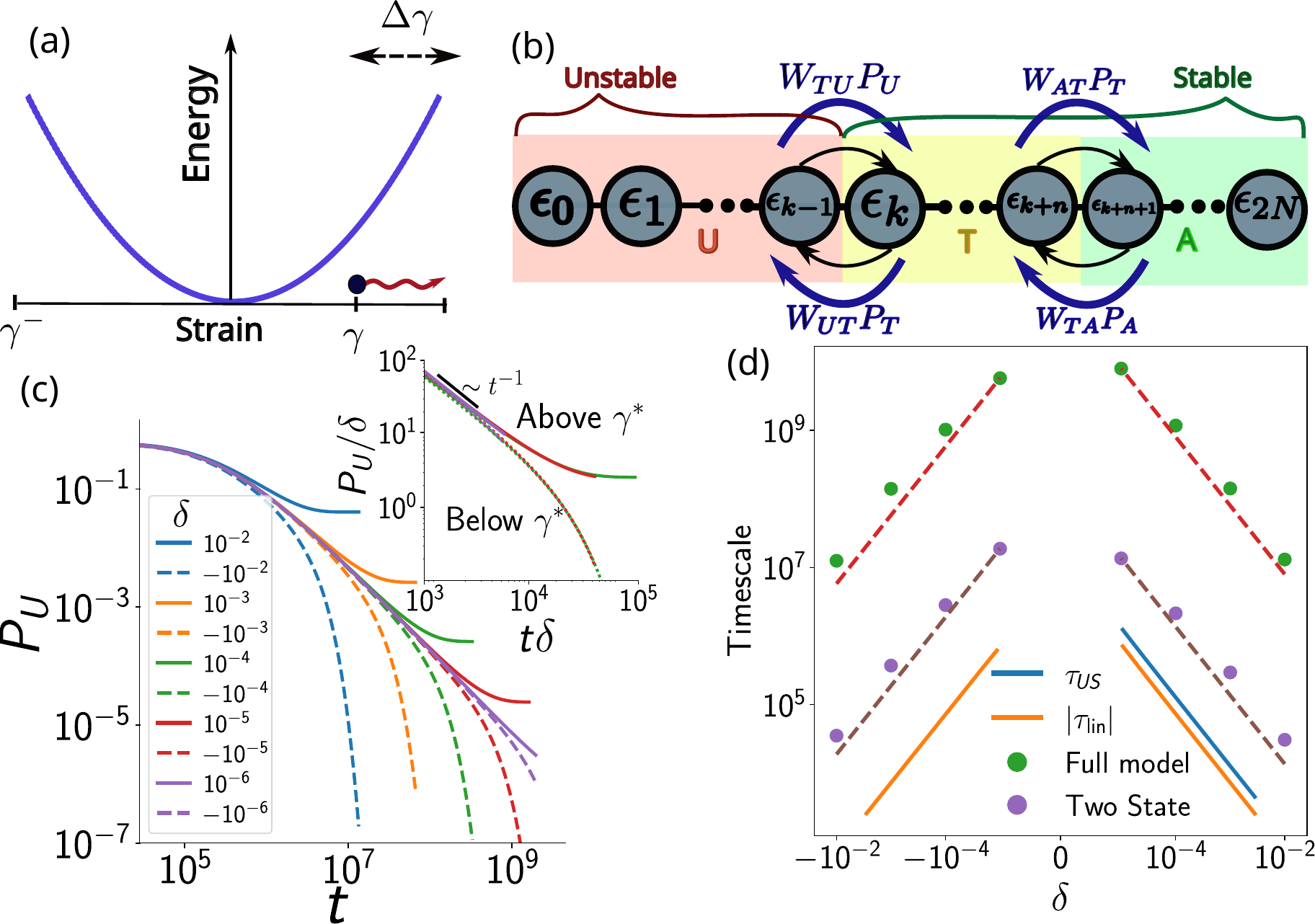}
    \caption{(a) Schematic diagram showing the energy of a state, $ = \epsilon$ at zero strain, and its variation with strain. The state becomes unstable when the strain energy $\gamma^2 = \epsilon$. Also shown is the strain interval $\Delta \gamma$, in excess of the imposed strain,  that needs to be traversed in order to make a transition to another state, which is envisaged to occur through a diffusion in strain. (b) Schematic diagram 
    of the full model, showing the transition rates between microstates, along with a coarse-graining into the stable 
    (absorbing (A) and threshold (T)) and unstable (U) states. The strain of amplitude $\gamma$ destabilizes all states below $k$. (c) Dynamics of $P_U$ near the critical strain amplitude $\gamma^*$ from numerical simulation of the full model, for strain $\gamma=\gamma^*(1+ \delta)$. At $\gamma^*$, $P_U\sim t^{-1}$. We see that $P_U(t)/\delta$ above and below $\gamma^*$ follows two scaling curves as shown in the inset. (d) Divergence of timescales about $\gamma^*$. $\tau_{US}=1/W_{US}$ is obtained from the two-state coarse graining; $\tau_\mathrm{lin}$ is determined by linearising \eqref{eq:two_state_master_equation}. We also show the timescales numerically obtained from the full model and the two-state model; all timescales diverge at $\gamma^*$ with exponent -1.}
    \label{fig:schematic_diffusion_parabola}
\end{figure}

As illustrated in Fig.~\ref{fig:schematic_diffusion_parabola}(a), for a state with energy $\epsilon$, with the application of shear $\gamma$, an additional strain interval $\Delta \gamma = \sqrt{\epsilon} - \gamma $, or the corresponding energy $\epsilon - \gamma^2$, must be overcome for it to yield. As discussed in \cite{mungan_metastability_2021}, such activation may be envisaged to arise from thermal noise or through mechanical noise. Escape rates corresponding to thermal activation were calculated in \cite{mungan_metastability_2021}, by mapping all states into two coarse-grained states, unstable (U) and stable (S) states (with probabilities $P_U$ and $P_S$), as illustrated in Fig. \ref{fig:schematic_diffusion_parabola}(b). It was shown that a yield value of $\gamma$ could be precisely identified across which the probability of being in stable (or frozen) states rapidly vanishes, in the steady state.

The propagation of {\it Eshelby} stresses accompanying localised plastic rearrangements has been extensively studied, and forms a basic ingredient of elasto-plastic models of the rheology of amorphous solids \cite{hebraud_mode-coupling_1998, agoritsas_relevance_2015, nicolas_deformation_2018, parley_mean-field_2022, liu_fate_2022, khirallah_yielding_2021}. Here, we consider a mean-field version of such an elastoplastic model where the plastic activity elsewhere in the system is treated as contributing random stress or strain increments at a given mesoscopic block, leading to diffusive dynamics of the strain of that block until the stability limit is reached. The strain diffusion coefficient is thus taken to be proportional to the probability of a mesoscopic block being in an unstable state: 
 \begin{equation}
     D(t)=\alpha P_U(t)
 \end{equation}
where the time dependence is explicitly indicated and $\alpha$ is a coupling constant that controls the strength of the interactions and is directly related to the elastic stress propagator \cite{agoritsas_relevance_2015, bocquet_kinetic_2009, parleyAgingMeanField2020}. An estimate of the additional time required to diffusively traverse $\Delta \gamma = \sqrt{\epsilon} - \gamma $ can be written as $ \frac{(\sqrt{\epsilon}-\gamma)^2}{2D}$. 
The time required to make a transition out of a state can then be written as 
\begin{equation}
    \tau_j=\begin{cases}
        \tau_{\mathrm{0}} & j \leq k,~k=2N\gamma^2 \\
        \tau_{0}\left(1+\frac{(\sqrt{\epsilon_j}-\gamma)^2}{2D}\right) & j > k. ~\label{eq:waiting_times}
    \end{cases}
\end{equation}

With these waiting times within the set of stable states, 

we choose as initial condition(s)
\begin{equation}
P_j(t=0)\propto \rho(\epsilon_j) e^{\beta \epsilon_j} 
\end{equation}
where $\beta$ parametrises the degree of annealing in the system. On doing so, we find that for some given value of $\alpha$, below a critical strain amplitude $\gamma^*$, the system reaches one of several steady states, depending on the initial state, for which $P_U=0$.

We refer to such states as `\textit{frozen}' (system) 

states. Above $\gamma^*$, we have additionally a steady state distribution with $P_U\neq 0$, which is the `\textit{fluidized}' state . We identify the value of $\gamma$ where such a steady state first appears as the common yield point $\gamma^*$. Above $\gamma^*$, whether the system will be in the frozen or fluidised state depends on its initial level of annealing.

We next write a coarse-grained description of the dynamics, following \cite{mungan_metastability_2021}, in terms of the variables $P_U$ and $P_{S}$, describing the fraction of mesoscopic blocks in unstable and stable states, respectively (cf.~Fig.~\ref{fig:schematic_diffusion_parabola}(b). This results in an effective dynamical system governed by
\begin{equation}
     \frac{\mathrm{d}}{\mathrm{d}t}\begin{pmatrix}P_U(t)\\
     P_S(t)
     \end{pmatrix}=\begin{pmatrix}
         -W_{SU}&W_{US}\\
         W_{SU}&-W_{US}
     \end{pmatrix}\begin{pmatrix}P_U(t)\\
     P_S(t)
     \end{pmatrix}.\label{eq:two_state_master_equation}
\end{equation}
where $W_{SU}$ depends on $\gamma$ (as calculated in {\it e.g.}~\cite{mungan_metastability_2021}) while $W_{US}$ depends, in addition, on the time dependent value of $P_U$ according to

(see SM \cite{SMref})

\begin{equation}
W_{US}= \frac{q_{k} \, p_{k-1,k}}{\tau_0} \frac{\alpha P_U}{a+\alpha P_U(1-b)} 
\end{equation}
where $q_k$ is the invariant distribution at the threshold state $\epsilon_k$ arising from the Ehrenfest transition rule and $b(\gamma)= \sum_{j = 0}^{k-1} q_j$ and $a(\gamma)=\sum_{j = k}^{2N} \, q_j (\sqrt{\epsilon_j}-\gamma)^2 /2$. The $\gamma$ dependences in $a$ and $b$ arise via the dependence on $k=2N\gamma^2$.

As $W_{US}$ is a function of $P_U$, the coarse-grained master equation \eqref{eq:two_state_master_equation} becomes nonlinear. Performing the fixed point analysis of the equation, we find that below a critical $\gamma^*$, only one fixed point corresponding to $P_U=0$ exists (frozen state), whereas for $\gamma>\gamma^*$, the fixed point $P_U=0$ becomes unstable and a fixed point with non-zero $P_U$ emerges (fluidized state), thus capturing the dynamical phase transition of the full model.

\paragraph*{Divergence of timescales:}
From the fixed point analysis, we find that below $\gamma^*$, the steady-state probabilities will be $P_U^*=0$ and $P_S^*=1$. Now, the steady-state probabilities and the two rates $W_{US}$ and $W_{SU}$ are related by the relation 

\begin{align}\label{steady-relations}
    P_U^*=\frac{W_{US}}{W_{US}+W_{SU}},\qquad P_S^*=\frac{W_{SU}}{W_{US}+W_{SU}}.
\end{align}
From these expressions, we see that the steady-state value of escape time $1/W_{US}$ will diverge on approaching $\gamma^*$ value from the right because $P_U^*(\gamma^*)=0$ and $W_{SU}$ is finite for all values of $\gamma$.

We can evaluate the divergence of the exponent by inverting \eqref{steady-relations} and expressing $W_{US}(P_U^*)$ and Taylor expanding $P_U^*$ about $\gamma^*$ as follows:
{\small
\begin{align}
    W_{US}(\gamma)=W_{SU}(\gamma)\frac{P_U^{*}(\gamma)}{1-P_U^*(\gamma)}\approx W_{SU}(\gamma)\frac{P_U^{*'}(\gamma^*)(\gamma-\gamma^*)}{1-P_U^{*'}(\gamma^*)(\gamma-\gamma^*)}
\end{align}
}
Given that $P_U^{*'}(\gamma^*)$ is finite, the escape time $\tau_{US}=1/W_{US}$ diverges as 
\begin{equation}\label{powerdiv}
    \tau_{US}(\gamma)\sim (\gamma-\gamma^*)^{-1},\qquad\gamma>\gamma^*.
\end{equation}
The divergence of timescale near $\gamma^*$ is also verified from numerical simulations of the full model as seen in Figure \ref{fig:schematic_diffusion_parabola} (c) and (d).
This divergence in escape time is reminiscent of the fatigue limit \cite{Basquin} where the number of cycles required for the system to fail below a certain stress/strain threshold diverges.

{\it Generalised Master Equation with mechanical noise:} 
While the two-state coarse-graining model above captures the dynamical (steady-state) fluidization transition, it can be shown to always lead to monotonic time-dependences of e.g.\ $P_U(t)$, being thus unable to describe ``fatigue failure dynamics'' where $P_U(t)$ first decays to small values but then eventually grows again, leading to fluidization. The general features of this dynamics can actually be studied for a more general model class, described by a master equation where the diffusivity arising from the mechanical noise is proportional to the occupation probability of being in the unstable states. We incorporate this effect into a generic discrete-state model with $N_u$ unstable states (block states that get destabilised within a shear cycle) and $N_s$ stable states. 
We will first show that a minimal model with $N_u = 1$ and $N_s = 2$ suffices to capture the full fatigue failure dynamics and then discuss such an effective model extracted by coarse-graining the Ehrenfest model with plasctic activity. We thus start out with a description of the general $(N_u,N_s)$ model class. 

We assume that the stable blocks yield due to the effects of noise and therefore have transition rates that vanish linearly as the occupation of unstable states goes to zero, as can be seen from the behaviour of $1/\tau_j$ in (\ref{eq:waiting_times}) for $D\to 0$.
Retaining only this linear dependence and assuming that, $D(t)=\alpha Y(t)$, the master equation governing the dynamics of such a system is 
\begin{align}\label{masterHL-main}
    \Dot{\mathbf{P}}_U&= \mathbf{W}_{UU}\mathbf{P}_U+\mathbf{W}_{US}\mathbf{P}_S\\
    \Dot{\mathbf{P}}_S&= \mathbf{W}_{SU}\mathbf{P}_U+\mathbf{W}_{SS}\mathbf{P}_S.
    \label{masterHL-main2}
\end{align}
where $\mathbf{W}_{US}=\alpha Y(t) \mathbf{W}_{US}'$ and $\mathbf{W}_{SS}=\alpha Y(t) \mathbf{W}_{SS}'$. Here $Y(t)$ is a measure of plastic activity in the system that could be taken as e.g.\ total occupation of unstable states as before, $Y=P_U=\mathbf{e}^T_U \mathbf{P}_U$ with $\mathbf{e}_U$ the indicator vector for unstable states ($=1$ for $j\in U$, $=0$ otherwise). Another plausible choice would be to set $Y(t)$ equal to the yielding rate in the unstable blocks,
\begin{equation}
    Y=-\mathbf{e}_U^T\mathbf{W}_{UU}\mathbf{P}_U.
\end{equation}
The scaling results for fatigue failure dynamics (i.e.\ small $P_U$) that we will derive are insensitive to the specific choice so we write in general $Y=\mathbf{g}^T \mathbf{P}_U$.
A fluidized state is then defined as before as one with $Y>0$, a frozen state as one with $\mathbf{P}_U=0$. Fatigue failure dynamics arises from the approach to {\em critically} frozen states, obtained in the limit $\mathbf{P}_U\to 0$ such that $\Dot{\mathbf{P}}_U$ also vanishes. More concretely, a critical frozen state is defined as
\begin{equation}
    \lim_{\mathbf{P}_U\to 0}\frac{\Dot{\mathbf{P}}_U}{|\Dot{\mathbf{P}}_S|}=0.
\end{equation}
Using \eqref{masterHL-main}, this yields the $N_u$ equations 
\begin{equation}
\mathbf{W}_{UU}\boldsymbol{\pi}_U+\alpha\mathbf{W}_{US}' \mathbf{P}_{S}=0
\end{equation}
where $\boldsymbol{\pi}_U=\lim_{\mathbf{P}_U\to 0} \mathbf{P}_U/Y=\lim_{\mathbf{P}_U\to 0} \mathbf{P}_U/(\mathbf{g}^T \mathbf{P}_U)$. Together with the normalizations 
\begin{equation}
    \mathbf{g}^T \boldsymbol{\pi}_U=1, \qquad 
    \mathbf{e}^T\mathbf{P}_S=1.
\end{equation}
one thus has $N_u +2$ equations for $N_u +N_s$ unknowns. Therefore, the solution space of critically frozen states must be $N_s -2$ dimensional. Thus, for $N_u=1$ and $N_s =2$, we will have a single critical frozen state. 

For simplicity, we therefore restrict ourselves to a discussion the case with $N_u=1$ and $N_s=2$, which turns out to be enough to study the salient features of the model, with the specific choice
$Y(t) = P_U(t)$. The master equation then takes the form

\begin{align}
    \Dot{P}_U&=P_U(-W_{TU}+ W_{UT}'P_T)\label{geneq1}\\
    \Dot{P}_T&=P_U(W_{TU}-(W_{UT}'+W_{AT}')P_T + W_{TA}'P_A)\label{geneq2}\\
    \Dot{P}_A&=P_U( W_{AT}'P_T -  W_{TA}'P_A)\label{geneq3}
\end{align}
where $P_U,~P_T,~P_A$ are the probability of being in the unstable and two stable states, denoted here T (transition) for close to being unstable and A (absorbing) for more stable. It turns out that this highly simplified model is already enough to predict fatigue failure dynamics in the form of non-monotonicity in the dynamics of $P_U$.

\paragraph*{Ehrenfest three state model:}
As stated earlier, a model of the form (\ref{geneq1}--\ref{geneq3}) can in fact be derived by coarse-graining the Ehrenfest model into three (rather than two) macrostates, defined explicitly as containing
(a) unstable states with $\epsilon\in(0,\gamma^2)$, (b) threshold states with $\epsilon\in(\gamma^2,\gamma^2+\lambda)$ where $\lambda$ parameterizes the number of local states deemed to be threshold states and could be used to optimize the coarse-graining, and (c) absorbing states with $\epsilon\in(\gamma^2+\lambda,1)$. 

\begin{figure}[!htbp]
    \centering
    \includegraphics[scale=0.2]{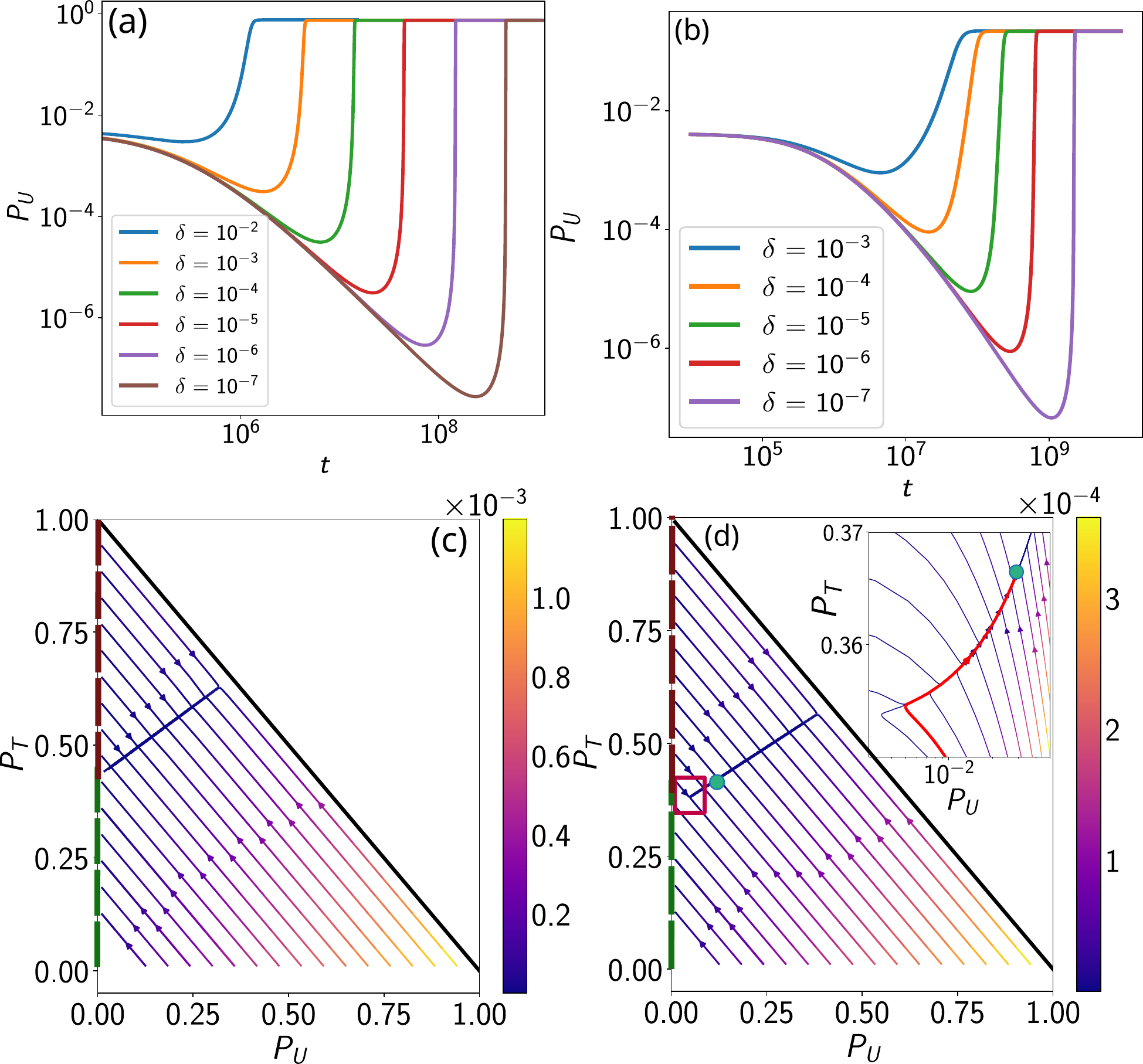}
    \caption{(a) Non-monotonic time-dependence of $P_U$ for a well-annealed sample with $\beta=40$ near critical strain amplitude $\gamma_c$ for different values of strain amplitude $\gamma=\gamma_c(1+\delta)$. The degree of annealing of the sample is $\beta=40$ and $\gamma_c =1.3273054\gamma^*$. (b) Non-monotonic trajectories from coarse-grained three state model starting from the initial condition marked with green in (c) for $\gamma>\gamma^*$. (c) and (d) Flow curves showing non-monotonic trajectories in the coarse-grained three state model for $\gamma<\gamma^*$ and $\gamma>\gamma^*$ respectively. For $\gamma<\gamma^*$, all states flow to $P_U=0$ no matter the initial conditions. At $\gamma\geq\gamma^*$, a new fixed point emerges marked by the green dot. The red and the green dash at $P_U=0$ shows the region of fixed point which is unstable and stable respectively. Inset: Zoomed-in region of Fig 2(d) marked in red square, showing the non-monotonic dynamics which gives rise to the non-monotonic fatigue failure behaviour.}
    \label{fig:non-mono}
\end{figure}
A local equilibrium assumption within each macrostate then gives explicit expressions for the rates in (\ref{geneq1}--\ref{geneq3}), including their dependence on the strain amplitude. Details are provided in a Supplemental Material \cite{SMref}.

Figure \ref{fig:non-mono} (a) and (b) respectively  show the resulting time evolution of $P_U$ for a well annealed sample in the full Ehrenfest model ( with $\beta=40$) and the coarse-grained Ehrenfest model respectively. Figure \ref{fig:non-mono} (c) and (d) shows flow curves of $P_U$ vs $P_T$ for strain $\gamma<\gamma^*$ and $\gamma>\gamma^*$ . In the former case, $P_U$ converges to zero for any initial condition whereas in the latter, there are initial conditions in the flow curves that generically show non-monotonic trajectories of $P_U(t)$. Interestingly, we also see trajectories for very well annealed samples (low $P_U$ and $P_T$) which converge to $P_U=0$. Note that the location of the boundary that separates these two types of different asymptotic states will depend on the actual value of $\gamma > \gamma^*$. The the range of initial configurations converging to a frozen state $P_U=0$ get progressively smaller with larger $\gamma$ values, becoming more and more confined to the region $P_U \approx P_T \approx 0$ and $P_A \approx 1$ of well annealed states. Thus, suppose for a given well annealed state (low value of $P_U$), and a given $\gamma > \gamma^*$, the system asymptotically evolves into a frozen state with $P_U = 0$. As $\gamma$ is increased, there will be a critical value of $\gamma_c > \gamma^*$, beyond which the asymptotic state is not frozen any more.  Moreover, this $\gamma_c$ will depend on the initial degree of annealing of the sample, being larger the better annealed the sample is. This latter result is consistent with the observation that well-annealed samples exhibit a higher yield strain amplitude $\gamma_c$ \cite{bhaumik_role_2021}.

\begin{figure}[!htbp]
    \centering
    \includegraphics[scale=0.24]{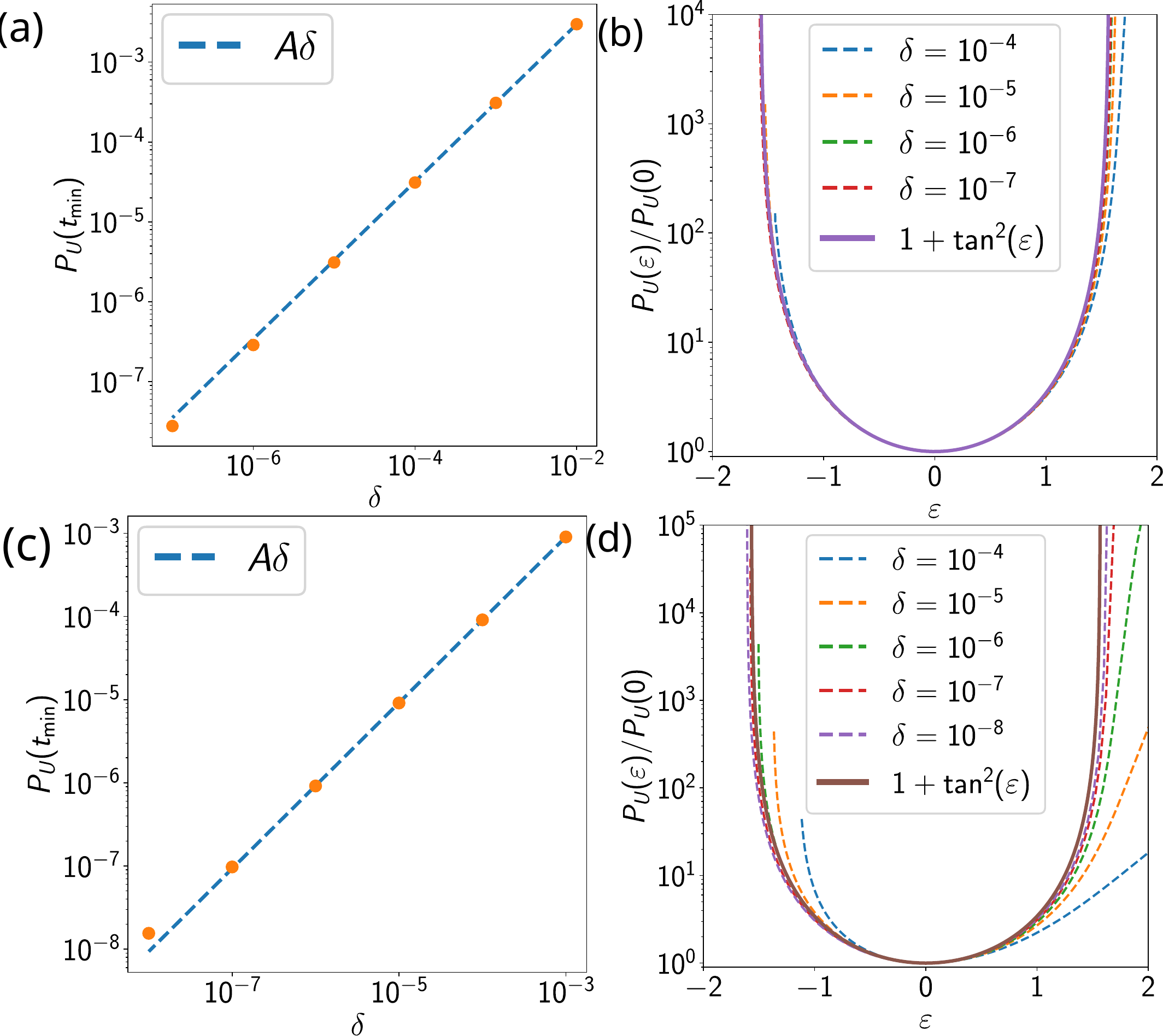}
    \caption{(a) Scaling of $P_U(t_\mathrm{min})$ for strain $\gamma=\gamma_c(1+\delta)$ for a well-annealed sample near critical strain amplitude $\gamma_c$. (b) Scaling collapse to the predicted $\tan^2$-form near the critically frozen state, with $\varepsilon=a (t-t_\mathrm{min}) \sqrt{P_U(t_\mathrm{min})}$ and $a$ a constant chosen to fit the scaling function. (c) Scaling of $P_U(t_\mathrm{min})$ with $\delta$ for the three state coarse-grained model for the initial condition given in Fig \ref{fig:non-mono}(c) above $\gamma_c$. (d) Corresponding scaling collapse.}
    \label{fig:fatigue}
\end{figure}
For $\gamma$ close to and above $\gamma_c$, when we numerically evolve the above equations near a critical frozen state, we find that $P_U$ decays to zero in a power-law manner. Indeed we can show analytically that close to the critical frozen state, the dynamics of $P_U$ can be expressed in a scaling form given by
\begin{equation}
    P_U=\delta \gamma \left(1+\tan^2 {\left(a\sqrt{\delta \gamma} ~\delta t \right)}\right)
    \label{p1_scaling}
\end{equation}
where $\delta \gamma =\gamma-\gamma_c$, $\gamma_c$ being the critical strain amplitude for the well-annealed sample and $\delta t=t-t_\mathrm{min}$. This is demonstrated for well-annealed samples in Figure \ref{fig:fatigue} (b) and (d) for the full model and the three-state coarse-grained version, respectively. Figure \ref{fig:fatigue} (a) and (c) show that the scaling of $P_U$ with $\delta \gamma$ also hold very well. The details of the calculation are given in the Supplementary Material \cite{SMref}. 

Remarkably, the scaling (\ref{p1_scaling}) can in fact be derived in full generality for the model defined by (\ref{masterHL-main},\ref{masterHL-main2}), with the analysis involving a reduction to a slow manifold within the space of unstable states.

From ~Eq. \eqref{p1_scaling}, one sees that in order for $P_U$ to have a value of $O(1)$, much bigger than its minimum value of $O(\delta\gamma)$, the $\tan^2$ needs to reach a correspondingly large value, and the time required $|\delta t|\sim \delta\gamma^{-1/2}$. The relaxation from some initial condition $P_U(0)=O(1)$ to $P_U(t_{\rm min})$ therefore takes a time $t_{\rm min}$ of this order. The time to re-fluidize after the fatigue failure minimum will have the same scaling, and thus the total time to fluidize near $\gamma_c$ will also scale as $\delta\gamma^{-1/2}$. Note that for $\gamma_c - \gamma^*$ sufficiently large, this time scale will dominate the approach to the asymptotic steady-state governed by the stable fixed point.

\paragraph*{Discussion:} We have investigated the yielding behaviour under cyclic shear employing a generalization of a model first studied in \cite{mungan_metastability_2021}, and including the effect of mechanical noise induced by interactions between mesoscopic blocks, in a mean field setting. States of local blocks that have large enough (yield) energies to be stable in the absence of any interaction with other mesoscopic blocks can be destabilised by the mechanical noise from other blocks. Representing this effect in a self-consistent mean-field manner, we find that the resulting model shows a genuine dynamical transition at a critical yield strain amplitude, accompanied by a power law divergence of timescales on either side of the transition. The non-trivial, non-monotonic, dynamics of the system leading to failure is correctly captured by a coarse-grained description of the system, composed of three distinct classes of states: unstable, threshold and absorbing states. Close to the transition, the system exhibits an initial regime of annealing, with decreasing plastic activity, following by an upturn leading to failure.  Such behaviour cannot be captured by a two state description of the system investigated earlier. Interestingly, the dynamics under three state coarse-graining exhibits an accumulation of probability of the threshold states, leading up to failure. The role of such accumulation will be interesting to study in future investigations. 
\bibliography{main}
\bibliographystyle{apsrev4-2}
\end{document}